\documentclass[11pt]{article}
\usepackage{jcappub}
\usepackage{amsmath}
\usepackage{amssymb}
\usepackage{accents}
\usepackage{subfigure}

\def\dual#1{\accentset{\boldsymbol{\neg}\vspace{-0.2ex}}{#1}}

\newcommand{\be}{\begin{equation}}
\newcommand{\ee}{\end{equation}}
\newcommand{\bea}{\begin{eqnarray}}
\newcommand{\eea}{\end{eqnarray}}
\newcommand{\bp}{\begin{pmatrix}}
\newcommand{\ep}{\end{pmatrix}}
\newcommand{\mh}{\mathcal{H}}

\newcommand{\ml}{\mathcal{L}}
\newcommand{\lsp}{\left(}
\newcommand{\rsp}{\right)}
\newcommand{\lsb}{\left[}
\newcommand{\rsb}{\right]}

\newcommand{\nn}{\nonumber}
\newcommand{\vp}{\varphi}
\newcommand{\ba}{\begin{appendix}}
\newcommand{\ea}{\end{appendix}}
\newcommand{\pd}{\partial}

\newcommand{\te}{\tilde{\eta}}

\title{Super-inflation and generation of first order vector perturbations in ELKO}

\author{Abhishek Basak}
\author{and S. Shankaranarayanan}

\affiliation{School of Physics, Indian Institute of Science Education and
Research Thiruvananthapuram (IISER-TVM), Trivandrum 695016, India}

\emailAdd{abhishek@iisertvm.ac.in}
\emailAdd{shanki@iisertvm.ac.in}

\abstract{In this work we construct  a model where the first order vector
perturbations can be generated during inflationary expansion. For the
non-standard spinors, known as ELKO, we show that the components
of the first order perturbed energy-momentum tensor of the ELKO is non-zero
for pure vector part of the metric perturbation ($B_{i}$). 
We show that
vector perturbations do not decay in the super-Hubble scale and for a
specific super-inflation background model we show that the vector
perturbations are nearly scale invariant, while its amplitude is 
smaller than the primordial scalar perturbations. We also 
comment on the generation of vorticity.}
 
\begin{document}

\maketitle

\section{Introduction} 
Inflationary paradigm has been highly successful in explaining the observed
Universe. In the standard inflationary scenario, inflation is driven by a
slowly rolling scalar field through its potential. The cosmological
perturbation theory during inflation has predicted the CMB
observations quite successfully, for example density perturbation or scalar
perturbations in the first order \cite{Kodama:1985bj,Mukhanov:1992ab}. 

However, generation of primordial seed magnetic field is still unresolved.
There are various mechanisms to generate primordial magnetic field
\cite{Turner:1987bw, Grasso:2000wj, Widrow:2002ud}.
One mechanism for the generation of primordial magnetic field is to
generate vorticity during inflation which can be sourced by primordial
vector modes. 
However, it has been observed that the first order perturbation theory can
not generate growing vector perturbations. The reason is easily understood
as the non-diagonal 
component of perturbed energy-momentum tensor do not contain
any vector modes. In the absence of anisotropic stress, the vector modes
decay quickly as the Universe expands \cite{Kodama:1985bj,Durrer:1993db}.
To avoid this problem of first order vector modes attempts have been made to
generate vector modes in the collapsing Universe during the contracting
phase of the cyclic models of the Universe \cite{Steinhardt:2001st}. It has
been shown that during this contracting phase the vector modes indeed grow,
but this growth cannot be stopped which may finally lead to breakdown of
the perturbation theory \cite{Battefeld:2004cd}. 

Unlike the first order perturbation, it has been observed
that in the case of second order perturbation theory the vector modes can
be sourced and vorticity can be generated \cite{Christopherson:2009bt,
Lu:2008ju, Christopherson:2010dw, Christopherson:2014eoa} 
even in the standard scalar field driven inflationary theory. 

In this work we consider the inflationary scenario driven by non-standard
spinors also known as $ELKO$ \cite{Ahluwalia:2004ab, Ahluwalia:2004sz,
Ahluwalia:2013uxa, daRocha:2008we}. 
These kind of spinors have mass dimension one and
follows Klein-Gordon equation instead of Dirac equation.
It has been shown that this kind
of spinors can drive inflation and lead to scalar power spectrum
consistent with the observed data \cite{Gredat:2008qf, Boehmer:2010ma,
Basak:2011wp}.
Here we show explicitly that ELKO driven inflation can generate growing
vector modes 
even in the first order. 
We show that, unlike the scalar field inflation, the non-diagonal
components of the stress-tensor (specifically ($\eta,i$) component, where
$\eta$ is the conformal time defined later and $i$ the spatial index) is
non-zero corresponding to the pure vector modes of the metric perturbation.
Rewriting the non-diagonal components, we show that the vector
perturbations, like the scalar perturbations, satisfy second order
differential equation.
In order to make a definite prediction, we assume that the
background field leads to super-inflation, i.e. $\dot H>0$
\cite{Gunzig:2000kk,Mulryne:2006cz,Biswas:2013dry,Liu:2013iha}, 
where $H$ is the Hubble
parameter.
The super-inflationary phase requires new physics which, in our case, is
provided by ELKO.
We have shown that the vector modes can be frozen in
the super-Hubble scale and is scale invariant. However, the amplitude is
small compared to the scalar perturbation. The suppression
factor of the amplitudes of the vector modes is $\exp\lsp-\frac{9}{8}\Delta
N\rsp$, where
$\Delta N$ is the number of e-foldings necessary for super-inflation. As in
the
case of ELKO driven inflationary theories, the spectral index of the scalar
modes of perturbations depends on the slow-roll parameter
$\epsilon=-\frac{\dot H}{H^{2}}$, 
super-inflationary phase $\lsp\dot H>0\rsp$ may produce blue
tilt in the spectral index of the scalar perturbations \cite{Basak:2011wp}.
Hence, the 
super-inflationary phase precedes the standard inflationary phase.
In order for the scalar perturbations from the ELKOS to be consistent 
with the CMB observations, the number of e-foldings in the super-inflation
phase can only be of the order unity. 
After the end of super-inflationary phase the standard
inflationary phase with $\lsp\dot H<0\rsp$ starts. 
 
In Sec. (2), we give the definitions of ELKO Lagrangian and the
energy-momentum tensor. In this section we further calculate the
non-diagonal component 
(specifically, ($\eta,i$)) of the perturbed energy-momentum tensor in the
linear order. In Sec. (3) we use the linear order perturbed Einstein
equation and using the perturbed energy-momentum tensor we calculate the
evolution equation of the pure vector modes $B_{i}$. At super-Hubble
scale, it is shown that the vector modes can be frozen in time. In Sec. (4)
we give the approximate background scaling solution, consistent with the
super-inflation. Sec. (5) contains the solutions of the vector modes in
the sub-Hubble and super-Hubble scale. Finally, in Sec. (6) we end with
conclusions and comment on the generation of vorticity in the first
order.

\section{Linear order perturbed ELKO energy-momentum tensor} 
In this work we are interested in the inflationary theory based on
non-standard spinor known as ELKO. ELKO are the eigenspinors of charge
conjugation operator. This kind of spinors are non-standard because, unlike
classical Dirac spinors in case of ELKO spinors 
$\lsp CPT\rsp^{-2}=-\mathbb{I}$. Other major difference between classical
Dirac spinors and ELKO is, Dirac spinors have mass dimension $\frac{3}{2}$
whereas ELKO spinors have mass dimension one. 
Therefore, these kind of spinors follow second order Klein-Gordon equation
instead of Dirac equation which is first order in time.
The energy-momentum tensor of ELKO can be written as \cite{Boehmer:2010ma}:
\be
T^{\mu \nu} = \dual{\lambda}
\overleftarrow{\nabla}^{(\mu}\overrightarrow{\nabla}^{\nu)}\lambda - 
g^{\mu \nu} \ml + F^{\mu \nu},
\label{em}
\ee
where $\lambda$ and $\dual{\lambda}$ are ELKO and its dual, respectively
and $\mu$ is 
the space-time index. 
The dual of ELKO is defined in the same spirit as is done in case of Dirac
spinors (Dirac adjoint in case of Dirac
spinors,$\bar\psi=\psi^{\dagger}\gamma^{0}$, $\gamma^{0}$ is the $0$-th
component of Dirac gamma matrix), 
such that $\lsp\dual{\lambda}\lambda\rsp$ becomes a space-time scalar. 
For a detailed discussion on the construction of ELKO and its dual, one
can look into \cite{Boehmer:2010ma}.
The ELKO Lagrangian ($\ml$) is  
\be
\mathcal{L} = \frac{1}{2} g_{\mu\nu}\dual{\lambda}\overleftarrow{\nabla}^{(\mu}
\overrightarrow{\nabla}^{\nu)}\lambda - V\lsp\dual{\lambda}\lambda\rsp
\ee
The exact form of the potential is arbitrary at this stage as it will be
clear in the following sections that it does not
not enter in the equation of motion of the vector modes directly.
The covariant derivative on ELKO and its dual are defined as:
\be
\dual{\lambda}\overleftarrow{\nabla} \equiv \partial^{\mu}\dual{\lambda} +
\dual{\lambda}\Gamma^{\mu},
\qquad
\overrightarrow{\nabla}\lambda \equiv \partial^{\mu}\lambda - \Gamma^{\mu}
\lambda.
\ee
Where $\Gamma_{\mu}$ is the spin-connection appearing because of
propagation of spinors in curved space-time. The expression of
$\Gamma_{\mu}$ is given by
\be
\Gamma_{\mu}=\frac{i}{4}\omega^{a b}_{\mu}f_{a b},
\ee
where $a$, $b$ are the spinor indices. 
$f^{a b}=\frac{i}{2}\lsb\gamma^{a},\gamma^{b}\rsb$ is the generator of the
Lorentz group and $\gamma^{a}$ is the Dirac gamma matrix. $\omega^{a
b}_{\mu}$ is defined as $\omega^{a b}_{\mu}=e^{a}_{\alpha}\lsp e^{\alpha
b}\rsp_{;\mu}$. Here $(;\mu)$ denotes the covariant derivative with respect
to $\mu$.
$e^{\mu}_{a}$ is the vierbien. The vierbiens are
related to the space-time metric $g^{\mu \nu}$ by the following relation:
\be
e^{\mu}_{a} e^{\nu}_{b} \eta^{a b}= g^{\mu \nu},
\ee
where $\eta^{a b}=\emph{diag.}\lsp 1, -1, -1, -1\rsp$ is the Minkowski metric.

$F^{\mu \nu}$ is the additional term that comes from
the variation of spin-connection, $\Gamma^{\mu}$, with respect to the 
metric $g^{\mu \nu}$. We work in the conformal time where the 
background metric is given as:
\be
g^{\lsp0\rsp}_{\mu \nu} = \bp a^{2}\lsp\eta\rsp & \mathbb{O} \\
                  \mathbb{O} & -a^{2}\lsp\eta\rsp \delta_{i j} 
              \ep.    
\ee
Here $\eta$ is the conformal time defined as $d\eta=dt/a$, $t$ is
the cosmic time and $a$ is the scale factor. 
The expression of $F^{\mu \nu}$ is given as
\be
F^{\mu \nu} = \frac{1}{2} \nabla_{\rho}J^{\mu \nu \rho},
\ee
where the expression of $J^{\mu \nu \rho}$ is:
\be
J^{\mu \nu \rho} = -\frac{i}{2}
\lsb\dual{\lambda}\overleftarrow\nabla^{( \mu}f^{\nu ) \rho}\lambda + 
\dual{\lambda} f^{\rho (\mu}\overrightarrow\nabla^{\nu )}\lambda\rsb.
\ee
$f^{\mu \nu}$ is given as $f^{\mu \nu} = e^{\mu}_{a} e^{\nu}_{b} f^{a
b}$. 

We prefer to work with the following ansatz for the back ground ELKO and
its dual
\be
\lambda = \vp\lsp\eta\rsp \xi, \qquad \dual{\lambda} = \vp\lsp\eta\rsp
\dual\xi,
\label{ans1}
\ee
where $\vp$ is real, 
$\xi$ and $\dual\xi$ are the two constant matrices with the property
\be
\dual\xi \xi = \mathbb{I},
\label{ans2}
\ee
such that $\dual{\lambda} \lambda = \vp\lsp\eta\rsp^{2}$.
In the above ansatz $\vp\lsp\eta\rsp$ is a background scalar quantity 
dependent on time only. The advantages of using the above 
ansatz are: (i) The components of
the energy-momentum tensor can be written in terms of one scalar field
$\lsp\vp\rsp$ instead of two spinors and (ii) it can be ensured that the 
theory does not have any negative energy or ghost modes \cite{Boehmer:2010ma}. 
Using (\ref{ans1}) and (\ref{ans2}) one can show that the ($\eta,\eta$)
component of the background energy-momentum tensor, which will be used
later, becomes:
\begin{equation}
T^{\eta \eta (0)} = a^{-4}
\lsb\frac{1}{2}\vp^{\prime
2}+\frac{3}{8}\mh^{2}\vp^{2}+a^{2}V\rsb, 
\label{emb1}
\end{equation}
where $'$ denotes derivative with respect to $\eta$ and $\mh$ is the Hubble
parameter defined as $\mh=a^{\prime}/a$.

The perturbed Einstein equation is given by:
\be
\delta G^{\mu}_{\nu} = 8\pi G \delta T^{\mu}_{\nu},
\label{pee}
\ee
For the vector modes in the metric perturbation we choose the Newtonian
gauge where the components of metric perturbation are given below:
\be
\delta g_{\eta i} = a^{2} B_{i},
\ee
where `$i$' is the spatial index -- $x$, $y$ and $z$. All other components
of the metric perturbation are equal to zero. Here $B_{i}$ is the 
divergence less vector mode i.e., $\partial^{i} B_{i} = 0$. 
The ($\eta,i$) component of the Einstein equation (\ref{pee}) is given by,
\be
\frac{1}{2} a^{-2} \Delta B_{i} = 8 \pi G \delta T^{\eta}_{i}, 
\label{pee:ei}
\ee
where $\Delta = \frac{\partial^{2}}{\partial x^{2}} +
\frac{\partial^{2}}{\partial y^{2}} + \frac{\partial^{2}}{\partial z^{2}}$.

It is important to note that in the standard scalar field driven
 inflation the right hand side of the equation (\ref{pee:ei}) vanishes. 
 Therefore, in case of the standard
 scalar field theory the solution of the vector modes $B_{i}$ are
identically zero in all scales. But here we will show that unlike the
standard canonical scalar field theory, in case of ELKO the expression of
$\delta T^{\eta}_{i}$ has the vector modes $B_{i}$. 

To calculate the perturbed energy momentum tensor we use the following
ansatz for the perturbed ELKO and its dual:
\be
\delta\lambda = \delta\vp\lsp\eta,\vec{x}\rsp\xi, \qquad 
\delta\dual{\lambda} = \delta\vp\lsp\eta,\vec{x}\rsp\dual\xi.
\ee
The expression of energy-momentum tensor (\ref{em}) shows that it is a sum
of three components: $T1^{\mu \nu} = \dual{\lambda}
\overleftarrow{\nabla}^{(\mu}\overrightarrow{\nabla}^{\nu)}\lambda$ ,
$T2^{\mu \nu} = g^{\mu \nu} \ml$ and $T3^{\mu \nu} = F^{\mu \nu}$. Here we
show the ($\eta,x$) component of the three terms of the perturbed 
energy-momentum tensor. The other components (($\eta,y$) and ($\eta,z$)) 
can be written accordingly.
The expression of $\delta T1^{\eta x}$, $\delta T2^{\eta x}$ and $\delta
T3^{\eta x}$ can be written as follows:

\bea
\delta T1^{\eta x} &=& a^{-4} 
\lsb B_{x}\vp^{\prime 2} - \vp^{\prime}\delta\vp_{,x} -
\frac{1}{8}\mh B_{x}^{\prime}\vp^{2}\rsb, \\
\delta T2^{\eta x} &=& a^{-4} \lsb\frac{1}{2}B_{x}\vp^{\prime 2} + 
\frac{3}{8}\mh^{2}B_{x}\vp^{2} - a^{2}B_{x}V\rsb, \\
\delta T3^{\eta x} &=& a^{-4} \lsb\frac{3}{4}\mh^{2}B_{x}\vp^{2} - 
\frac{1}{16}B_{x}^{\prime\prime}\vp^{2} - 
\frac{1}{8}B_{x}^{\prime}\vp\vp^{\prime} + 
\frac{1}{4}\mh\vp\delta\vp_{,x} +
\frac{1}{8}\Delta B_{x}\vp^{2}\rsb.
\eea
In the above expression $(_{,x})$ denotes the partial derivative with
respect to $x$.
Therefore, one can generalize ($\eta,i$) component of the
energy-momentum tensor for vector perturbation as:

\bea
\nn\delta T^{\eta i} &=& a^{-4} 
[ \frac{1}{2}B_{i}\vp^{\prime 2} - \vp^{\prime}\delta\vp_{,i} -
\frac{1}{8}\mh B_{i}^{\prime}\vp^{2} + 
\frac{3}{8}\mh^{2}B_{i}\vp^{2} + a^{2}B_{i}V -
\frac{1}{16}B_{i}^{\prime\prime}\vp^{2} - 
\frac{1}{8}B_{i}^{\prime}\vp\vp^{\prime} + \\
&&\frac{1}{4}\mh\vp\delta\vp_{,i} +
\frac{1}{8}\Delta B_{i}\vp^{2}]
\eea
The expression of energy-momentum tensor in the mixed form can be
calculated using the following relation:
\be
\delta T^{\mu}_{\nu} = \delta T^{\mu \sigma} g^{(0)}_{\sigma \nu} + 
                       T^{\mu \sigma (0)} \delta g_{\sigma \nu}.
\ee
Therefore, using the expression (\ref{emb1}) ($\eta,i$) component
of the energy-momentum tensor in the mixed form becomes:

\be
\delta T^{\eta}_{i} = a^{-2} 
\lsb \frac{1}{8}\mh B_{i}^{\prime}\vp^{2} +
\frac{1}{16}B_{i}^{\prime\prime}\vp^{2} + 
\frac{1}{8}B_{i}^{\prime}\vp\vp^{\prime} + 
\lsp\vp^{\prime} - \frac{1}{4}\mh\vp\rsp\delta\vp_{,i} -
\frac{1}{8}\Delta B_{i}\vp^{2}\rsb
\label{emp1}
\ee


\section{Evolution equation of $B_{i}$}
Substituting equation (\ref{emp1}) in the equation (\ref{pee:ei}) the 
expression of the Einstein equation can be written as:
\be
\Delta B_{i}\lsb 1+\frac{1}{4}\frac{\vp^{2}}{M^{2}_{pl}}\rsb = 
\frac{1}{M^{2}_{pl}} 
\lsb \frac{1}{4}\mh B_{i}^{\prime}\vp^{2} +
\frac{1}{8}B_{i}^{\prime\prime}\vp^{2} + 
\frac{1}{4}B_{i}^{\prime}\vp\vp^{\prime} + 
2\lsp\vp^{\prime} - \frac{1}{4}\mh\vp\rsp\delta\vp_{,i}\rsb,
\label{pee1}
\ee
where $M_{pl} = \frac{1}{\sqrt{8 \pi G}}$ is the reduced Planck Mass. 
The above equation can be rewritten as 
\be
B^{\prime\prime}_{i}+2\lsp\frac{\vp^{\prime}}{\vp}+\mh\rsp B^{\prime}_{i}-
\lsp2+\frac{8M^{2}_{pl}}{\vp^{2}}\rsp \Delta B_{i}+
16\lsp\vp^{\prime} -
\frac{1}{4}\mh\vp\rsp\frac{\delta\vp_{,i}}{\vp^{2}}=0,
\label{pee11}
\ee
In the Fourier mode the equation (\ref{pee1}) can be written as 
\be
B^{\prime\prime}_{i}+A_{1}B^{\prime}_{i}+A_{2}k^{2}B_{i}-
16ik_{i}\lsp\vp^{\prime} -
\frac{1}{4}\mh\vp\rsp\frac{\delta\vp}{\vp^{2}}=0,
\label{pee2}
\ee
where, $A_{1}=2\lsp\frac{\vp^{\prime}}{\vp}+\mh\rsp$ and
$A_{2}=2+\frac{8M^{2}_{pl}}{\vp^{2}}$. The Fourier modes are related to the
partial derivatives as $\pd_{i}\equiv-ik_{i}$. The last term of the above
equation acts as a source term. 
For
simplicity, one can set the last term in Eq.~($\ref{pee2}$) to vanish,
i.e., 
\be
\lsp\vp^{\prime} - \frac{1}{4}\mh\vp\rsp = 0.
\label{suincon}
\ee
Under this condition, the evolution equation (\ref{pee11}) simplifies to
\be
B^{\prime\prime}_{i}+A_{1}B^{\prime}_{i}-A_{2}\Delta B_{i}=0.
\label{pee3}
\ee
The above equation looks very similar to
the evolution equation of scalar perturbations during inflation. 
In the next section we show that the condition (\ref{suincon}) leads to
consistent background evolution.


\section{Background scaling solution}

Before proceeding with the power-spectrum calculation, we 
show that the condition (\ref{suincon}) leads to a consistent background 
evolution and that it leads to Super-inflation $(\dot{H} > 0)$.

The Klein-Gordon equation of the background field $\vp$ is given as
\cite{Boehmer:2010ma}
\be
\vp^{\prime\prime}+2\mh\vp^{\prime}-\frac{3}{4}\mh^{2}\vp+a^{2}V_{,\vp}=0,
\label{bck1}
\ee
where ($_{,\vp}$) denotes the derivative with respect to the background
field $\vp$. The modified Friedmann equations are given as follows:
\begin{eqnarray}
\mh^{2}&=& \frac{1}{1-\tilde{F}}\left[\frac{1}{3M^{2}_{\rm pl}}\left(\frac{\varphi^{\prime 2}}{2}+
                                              a^{2}V\right)\right],\label{bck2}\\
 \mh^{\prime}&=& \frac{1}{1-\tilde{F}}\left[\frac{1}{3M^{2}_{\rm pl}}\left(a^{2}V-\varphi^{\prime 2}\right)+
                          \mh\tilde{F}^{\prime}\right], \label{bck3}
\end{eqnarray}
where $\tilde{F}=\frac{\vp^{2}}{8M^{2}_{Pl}}$. To find the background
scaling solutions for power-law type of potential we choose the following
forms of scale factor, background field and potential:
\be
a\lsp\eta\rsp=A\lsp-\eta\rsp^{-q}, \qquad
\vp\lsp\eta\rsp=\vp_{0}\lsp-\eta\rsp^{p}, \qquad
V=V_{0}\vp^{\beta},
\label{scal1}
\ee
where $A$, $\vp_{0}$ and $V_{0}$ are some arbitrary constants which can be
expressed in terms of the exponents $-q$, $p$ and $\beta$ using the three
background equations. Keeping in mind the condition that
$\vp^{\prime}=\frac{1}{4}\mh\vp$ one can easily write $p=-nq$, where
$n=\frac{1}{4}$.
The equations (\ref{bck2}) and (\ref{bck3}) are qualitatively same. 
Substituting (\ref{scal1}) in the background equations (\ref{bck1}) and
(\ref{bck2}) one find the following relations between $\beta$ and $q$
respectively
\be
q=2/\lsp 2+n\beta-2n\rsp, \qquad q=2/\lsp 2+n\beta\rsp.
\ee
Therefore, one
can consistently solve the background equations for 
\be
\beta\gg2,\qquad 0<q\ll1.
\ee

Writing the background equation (\ref{bck3}) in cosmic time gives us:
\be
\dot H\lsp1-\pi G\vp^{2}\rsp = -4\pi G \dot\vp^{2} + 2\pi G H
\vp\dot\vp,
\label{bge1}
\ee
where $H=\mh/a$. From the above expression one can see that when
$\frac{\dot\vp}{H\vp}>\frac{1}{2}$ one gets the standard inflationary
theory with $\dot H<0$ and when $\frac{\dot\vp}{H\vp}<\frac{1}{2}$ one gets
the super-inflationary theory with $\dot H>0$.
Therefore, using the condition (\ref{suincon}) the
equation (\ref{bge1}) tells us that $\dot H>0$. One can also
show that for $0<q<1$, the expression of the scale factor
($a\lsp\eta\rsp=A\lsp-\eta\rsp^{-q}$) gives us $\dot H>0$. So, under
the condition $0<q<1$ one gets positive acceleration and at the same
time one can see that the scale factor grows in cosmic time.
This phase of evolution is known as super-inflation
\cite{Gunzig:2000kk,Mulryne:2006cz,Biswas:2013dry}. 


\section{Solutions in the sub-Hubble and super-Hubble scale}
Following \cite{Martin:2007ue, Subramanian:2009fu} the general solution of
(\ref{pee3}) can be written as:
\be
B_{i}= \int\varepsilon_{i r}\lsp\vec{k}\rsp\lsb b_{r}\lsp\vec{k}\rsp 
B\lsp\eta,k\rsp
e^{i\vec{k}.\vec{x}}+
b_{r}^{\dagger}\lsp\vec{k}\rsp 
B^{*}\lsp\eta,k\rsp
e^{-i\vec{k}.\vec{x}}
\rsb \frac{d^{3}k}{\lsp2\pi\rsp^{3/2}}.\label{pee31}
\ee
$b_{r}$ and $b_{r}^{\dagger}$ are the annihilation and creation operators
respectively.
Here $\varepsilon_{\mu r}\lsp\vec{k}\rsp=\lsp
0,\vec{\varepsilon}_{r}\lsp\vec{k}\rsp\rsp$ is the 
polarisation vector and $r=1,2,3$. $\vec{\varepsilon}_{1}\lsp\vec{k}\rsp$,
$\vec{\varepsilon}_{2}\lsp\vec{k}\rsp$ are the mutually orthogonal unit vectors
also orthogonal to $\vec{k}$ and $\vec{\varepsilon}_{3}\lsp\vec{k}\rsp$ is the
unit vector along the direction of $\vec{k}$. 
Here we have defined the polarisation vectors slightly differently than in the
references \cite{Martin:2007ue, Subramanian:2009fu} (check appendix
(\ref{app1}) for discussion). The advantage of doing so is
that, instead of redefining the scalar ($\bar{B}=aB$) \cite{Martin:2007ue,
Subramanian:2009fu}, in Fourier mode one can now directly write the
evolution equation (\ref{pee3}) in terms of the scalar quantity
$B\lsp\eta,k\rsp$ by replacing $\Delta$ with $-k^{2}$,
\be
B^{\prime\prime}+A_{1}B^{\prime}+A_{2}k^{2} B=0.
\label{pee32}
\ee

One can see that the coefficient of $k$, $\sqrt A_{2}$, in equation
 (\ref{pee32}) is not a constant in time. 
 One can remove the time dependent coefficient of $k^{2}$ by
  redefining the time
 parameter as $d\te=\sqrt{A_{2}}d\eta$ \cite{Kobayashi:2010cm}.
 Under this change of the time variable, equation (\ref{pee3}) can be
 written as 
 \be
B_{,\te\te}+\tilde{A_{1}} B_{,\te}+k^{2}B=0,
\label{pee7}
 \ee
where
$\tilde{A_{1}}=\frac{1}{2}\frac{A_{2,\te}}{A_{2}}+\frac{A_{1}}{\sqrt{A_{2}}}$.
Here, in terms of $\te$, $A_{1}$ becomes 
$A_{1}\lsp\te\rsp=2\sqrt{A_{2}}\lsp\frac{\vp_{,\te}}{\vp}+\tilde{\mh}\rsp$,
where $\tilde\mh=a_{,\te}/a$.
 One can eliminate $B_{,\te}$ term from equation (\ref{pee7}) by
 redefining $B\lsp\te,k\rsp= \mathcal{B}\lsp\te,k\rsp f\lsp\te\rsp$. 
 Substituting this form of $B$ in
equation (\ref{pee7}) one can eliminate $\mathcal{B}_{,\te}$ term by
setting its coefficient as zero, which finally gives us 
\bea
f=\exp\lsp-\frac{1}{2}\int\tilde{A_{1}}d\te\rsp,\\
\mathcal{B}_{, \te\te}+\lsb k^{2}-\lsp\frac{\tilde{A}_{1,\te}}{2}+
\frac{\tilde{A}^{2}_{1}}{4}\rsp\rsb \mathcal{B}=0.
\label{pee4}
\eea

From the equation (\ref{bck2}) one can see that $\vp^{2}<8M^{2}_{Pl}$, as
$\mh$ can not be imaginary. Thus, when $\frac{8M^{2}_{Pl}}{\vp^{2}}$  
is much larger than $2$, one can write
$\frac{A_{2,\te}}{A_{2}}\approx-2\frac{\vp_{,\te}}{\vp}$. Therefore, under
this condition the expression of $\tilde{A_{1}}$ becomes
$\tilde{A_{1}}\approx\frac{\vp_{,\te}}{\vp}+2\tilde{\mh}$.
Using the condition (\ref{suincon}) (which, in terms of $\te$, remains
unchanged) the factor $f$ becomes
$f=\exp\lsp-\frac{9}{8}\Delta N\rsp$. Here $\Delta N\approx\int{\tilde{\mh}
d\te}=\int{\mh d\eta}$ denotes the number of
e-folding required for super-inflation. This factor acts as a suppressing
factor. The larger the number of e-folding, the larger is the
suppressing factor.

Using the expression of $\tilde{A}_{1}$ and the condition (\ref{suincon})
in terms of $\te$ equation
(\ref{pee4}) can be expressed in terms of $\mh$ as
\be
\mathcal{B}_{, \te\te}+\lsb k^{2}-\lsp\frac{9}{8}\tilde{\mh}_{,\te}+
\frac{81}{64}\tilde{\mh}^{2}\rsp\rsb \mathcal{B}=0.
\label{pee5}
\ee
Using (\ref{scal1}) and the expression of $\sqrt{A_{2}}$ one can write the
expression of $\te$ in terms of $\eta$ as:
$-\te=\frac{\lsp-\eta\rsp^{1-2p}}{1-2p}$. Therefore, one can see that 
$\eta\rightarrow-\infty\implies\te\rightarrow-\infty$ and
$\eta\rightarrow 0\implies\te\rightarrow 0$, when $p\ll1$.
Thus, the expression of the scale factor in terms of $\te$ becomes
$a\lsp\te\rsp=\lsp-\te\rsp^{-m}$, where $m=\frac{q}{1-2p}$. Therefore, in
this case also $0<m\ll1$.
Finally, using the expression of $a(\te)$ the evolution
equation (\ref{pee5}) can be expressed in terms of Bessel differential
equation:
\be
\mathcal{B}_{,\te\te}+\lsb k^{2}-
\frac{1}{\te^{2}}\lsp\nu^{2}-\frac{1}{4}\rsp\rsb \mathcal{B}=0,
\label{pee6}
\ee
where
$\nu^{2}=
\frac{1}{4}\lsb1+\lsp\frac{153}{16}+\frac{9}{2}\tilde{\epsilon}\rsp m^{2}\rsb$. 
Where $\tilde{\epsilon}=1-\frac{\tilde{\mh}_{,\te}}{\tilde{\mh}}$ is the 
slow-roll parameter in the standard inflationary scenario. 
However, it should be
noted that in the super-inflationary phase $\tilde{\epsilon}$ can be $\sim
1$ or large. 
In the super-inflationary phase one can achieve acceleration without
$\tilde{\epsilon}$ requiring to be smaller than one. 

As $\nu$ is positive, the solution of the equation can be expressed in
terms of the Hankel function of the first and second kind
\be
\mathcal{B}_{k} = \sqrt{-\te}\lsb a_{1}\lsp
k\rsp H^{(1)}_{\nu}\lsp x\rsp+a_{2}\lsp
k\rsp H^{(2)}_{\nu}\lsp x\rsp\rsb,
\label{peesol}
\ee
where $x=-k\te$.

In the limit of $x\gg1$, the properties of the Hankel functions are following
\bea
\nn
H^{(1)}_{\nu}\lsp x\gg1\rsp &\approx& \sqrt{\frac{2}{\pi x}}
e^{ix-i\frac{\pi}{2}\lsp\nu+\frac{1}{2}\rsp},\\
H^{(2)}_{\nu}\lsp x\gg1\rsp &\approx& \sqrt{\frac{2}{\pi x}}
e^{-ix+i\frac{\pi}{2}\lsp\nu+\frac{1}{2}\rsp}. \label{hankel1}
\eea
Following the methods used in case of standard inflationary theory,
to match with the plane wave solution in the sub-Hubble scale 
-- $\mathcal{B}_{k}\sim\frac{e^{ix}}{\sqrt{k}}$ -- one can 
set $a_{2}=0$ and 
$a_{1}=\frac{\sqrt{\pi}}{2}e^{i\frac{\pi}{2}\lsp\nu+\frac{1}{2}\rsp}$.
The property of the Hankel function $H^{(1)}_{\nu}\lsp x\rsp$, in the limit
of $x\ll1$ becomes
\be
H^{(1)}_{\nu}\lsp x\ll1\rsp \approx \sqrt{\frac{2}{\pi}}
 e^{-i\frac{\pi}{2}}2^{\lsp\nu-\frac{3}{2}\rsp}
\frac{\Gamma\lsp\nu\rsp}{\Gamma\lsp\frac{3}{2}\rsp}x^{-\nu}.
\label{hankel2}
\ee
Thus, the solution of $\mathcal{B}_{k}$ in the super-Hubble
scale can be written as 
\be
\mid\mathcal{B}_{k}\mid\sim \sqrt{-\te}x^{-\nu}=
\frac{k}{\sqrt{k^3}}\lsp k\rsp^{\frac{1}{2}-\nu}
\lsp-\te\rsp^{\frac{1}{2}-\nu}.
\label{ui}
\ee
Using the fact that the solutions nearly remains unchanged after Hubble
crossing, one can use $k=\tilde{\mh}$ during Hubble crossing and the expression
(\ref{ui}) can be rewritten as
\be
\mid\mathcal{B}_{k}\mid\sim \frac{\tilde{\mh}}{\sqrt{k^3}}
\lsp k\rsp^{\frac{1}{2}-\nu}\lsp-\te\rsp^{\frac{1}{2}-\nu}.
\label{ui1}
\ee
Finally, the expression of $\lsp B\rsp$ can be written as:
\be
\mid B_{k}\mid\sim e^{-\frac{9}{8}\Delta N} 
\frac{\tilde{\mh}}{\sqrt{k^3}}
\lsp k\rsp^{\frac{1}{2}-\nu}
\lsp-\te\rsp^{\frac{1}{2}-\nu}.
\label{bi}
\ee
From the expression of $\nu$ one can identify the spectral index $n_{V}$
for vector modes as:
\be
n_{V}=\lsp\frac{153}{16}+\frac{9}{2}\tilde{\epsilon}\rsp m^{2}.
\ee
For a small value of $m$ one can approximate
$\nu\sim\frac{1}{2}$.
Therefore, from equation (\ref{bi}) one can see that in the super-Hubble
scale the vector modes will be nearly frozen and nearly scale independent
similar to the scalar perturbations. It is important to note that there are
no observational evidence for scale independent vector perturbation yet.
However, super-inflationary phase naturally provides the scale invariance.
One can identify that the term similar to
$\frac{\tilde{\mh}}{\sqrt{k^{3}}}$ also appears in the amplitude of the scalar
perturbations, for example one can see reference \cite{Riotto:2002yw}. 
But in case of vector modes the amplitude is suppressed by
the factor ($e^{-\frac{9}{8}\Delta N}$) compared to the scalar modes. 
Here, $\Delta N$
is the number of e-foldings required for super-inflation. 
The super-inflationary phase requires new physics which in our case is
provided by ELKO. 
Hence, the super-inflationary phase precedes the standard
inflationary phase 
and the number of e-folding is $\sim \mathcal{O}\lsp1\rsp$. The observation
of the vector modes in the CMB polarization can restrict the number of
e-foldings of super-inflation.

\section{Conclusion}
In this work we have calculated the perturbed energy-momentum tensor of
the ELKO. It has been shown that unlike the standard scalar field case, the
non-diagonal component (specifically, ($\eta,i$)) of the stress-energy tensor
is non-zero.
The same
component of the Einstein 
equation gives us the evolution equation of the vector modes ($B_{i}$)
which looks similar to the scalar perturbation equation. 
We have analytically obtained the vector perturbations for the background
evolution satisfied by 
$\lsp\vp^{\prime}=\frac{1}{4}\vp\mh\rsp$. We have shown that this condition
leads to super-inflation ($\dot H>0$). 

We have shown explicitly that the vector perturbations are nearly scale
invariant and frozen in the super-Hubble scales. However, the amplitude of
these perturbations are smaller
compared to the scalar perturbation.

In the super-Hubble scale 
one can look at the behaviour of the solution
of (\ref{pee2}) by setting $k\rightarrow0$. As 
$k=\sqrt{k^{2}_{x}+k^{2}_{y}+k^{2}_{z}}$, each $k_{i}$ will also be small,
i.e., $k_{i}\rightarrow0$. 
As, so far we have not observed any vector modes, one can presume that they
have smaller amplitude than the scalar perturbation $\delta\vp/\vp$.
Therefore, in general, one can not ignore the last term in (\ref{pee2}) in
the super-Hubble scale.
However, as $\delta\vp/\vp$ is restricted by the observation to be very
small, we presume that it is always possible to find a suitable $k_{i}$ for
which the last term in (\ref{pee2}) can also be smaller than the first two
terms. Under such conditions, in the super-Hubble scale,
equation (\ref{pee2}) can be simplified as
\be
B^{\prime\prime}_{i}+A_{1}B^{\prime}_{i}=0,
\label{peesim}
\ee
which tells us that for positive value of $A_{1}$, the solution for $B_{i}$
is frozen in the super-Hubble scale for a given initial condition, similar
to the scalar perturbation. However, the initial condition is given by the
solutions in the sub-Hubble scale, which can bring the $k$ dependence in
the full solutions of (\ref{pee2}).
In this work we have used the condition $\vp^{\prime}=\frac{1}{4}\vp\mh$ which
gives us super-inflation. However,
one can understand from equation (\ref{peesim}) that during
standard inflation, even when the above condition is violated, the vector modes
will be nearly constant in time in the super-Hubble
scale. This needs further investigation.

As, usually the vorticity in the perfect fluid follow nearly the similar
equation
as the vector modes (in this case it will be second order differential
equation in time with some additional terms), hence, the vorticity
can also be generated in 
the first order perturbation theory. With the generated vorticity one can
also look into the production of large-scale primordial magnetic field.
Once the {\em{Planck}} polarization
results are
published, the consequences of this kind of models of vector perturbations 
can be verified. For the observational possibilities one can look into the
Ref. \cite{Lewis:2004kg}.

We know that most of the
inflationary models with standard scalar fields can not produce pure vector
modes in the first order. 
Apart from producing scalar perturbations
consistent with observations, the inflationary theory driven by
non-standard spinors like ELKO can also produce pure vector modes in the
first order.
Therefore, the observation of vector modes in the CMB polarization can make
this kind of non-standard spinors a potential candidate for inflaton.


\section{Acknowledgments}

We thank T. R. Seshadri and Kandaswamy Subramanian  for useful discussions. 
The work is supported by Max Planck-India 
Partner Group on Gravity and Cosmology. SS is partially supported by
Ramanujan Fellowship of DST, India.
\appendix
\section{{Appendix: Polarisation vector in the curved space-time\label{app1}}}
Following the formalism in the quantum field theory (for example, one can
look at \cite{Mandl:2010qu}) in flat Minkowski space-time, the
orthonormality and completeness relations of the polarisation vectors
$\varepsilon^{\mu}_{r}\lsp\vec{k}\rsp$ can be written as
\be
\varepsilon_{r}\lsp\vec{k}\rsp\varepsilon_{s}\lsp\vec{k}\rsp=\varepsilon_{r
\mu}\lsp\vec{k}\rsp\varepsilon^{\mu}_{s}\lsp\vec{k}\rsp=-\zeta_{r}\delta_{r
s}, \qquad r=0,...,3,
\label{a1}
\ee
\be
\sum\limits_{r}\zeta_{r}\varepsilon^{\mu}_{r}\lsp\vec{k}\rsp\varepsilon^{\nu}
_{s}\lsp\vec{k}\rsp=-\eta^{\mu\nu},
\label{a2}
\ee
\be
\zeta_{0}=-1, \qquad \zeta_{1}=\zeta_{2}=\zeta_{3}=1.
\label{a3}
\ee
Here $\varepsilon_{r
\mu}\lsp\vec{k}\rsp=\eta_{\mu \nu}\varepsilon^{\nu}_{r}\lsp\vec{k}\rsp$.
The above formalism is used in the quantisation of the electromagnetic
field ($A^{\mu}$) in the Minkowski space-time.
One can choose the polarisation vector as
$\varepsilon^{\mu}_{r}\lsp\vec{k}\rsp=\lsp0,
\vec{\varepsilon}_{r}\lsp\vec{k}\rsp\rsp$, where $r=1,...,3$.
$\vec{\varepsilon}_{1}\lsp\vec{k}\rsp$,
$\vec{\varepsilon}_{2}\lsp\vec{k}\rsp$ are the mutually orthogonal unit vectors
also orthogonal to $\vec{k}$. $\vec{\varepsilon}_{3}\lsp\vec{k}\rsp$ is the
unit vector along the direction of $\vec{k}$. Therefore,
$\vec{\varepsilon}_{1}\lsp\vec{k}\rsp$,
$\vec{\varepsilon}_{2}\lsp\vec{k}\rsp$ are also orthogonal to
$\vec{\varepsilon}_{3}\lsp\vec{k}\rsp$.

To generalise the expressions (\ref{a1}), (\ref{a2}) and (\ref{a3}) in the
curved space-time, in reference \cite{Martin:2007ue, Subramanian:2009fu}
the authors have chosen to multiply $1/a$ with all components of the 
polarisation 
vector $\varepsilon^{\mu}_{r}\lsp\vec{k}\rsp$. Then one can replace
$\eta^{\mu\nu}$ with $g^{\mu\nu}$ in (\ref{a2}) and the expression
(\ref{a3}) remains unchanged. \emph{This formalism gives correct
solutions of the electro-magnetic vector potential $A_{i}$} as one can
always absorb the scale factor in the scalar term which
appears in the Fourier decomposition of $A_{i}$. However, this formalism
tells us that $\varepsilon^{\mu}_{r}\lsp\vec{k}\rsp$ is no longer a
function of only $\vec{k}$, it also becomes a function of time because of
the presence of the scale factor. 

Keeping in mind the above points, we propose a slightly different formalism
of polarisation vectors to quantize the vector fields of contravariant
and covariant form in the curved space-time.

\noindent (i) Vector fields of contravariant form $A^{\mu}$: The Fourier
decomposition is given as: 
\be
A^{\mu}= \int\varepsilon^{\mu}_{r}\lsp\vec{k}\rsp\lsb b_{r}\lsp\vec{k}\rsp 
A\lsp\eta,k\rsp
e^{i\vec{k}.\vec{x}}+
b_{r}^{\dagger}\lsp\vec{k}\rsp 
A^{*}\lsp\eta,k\rsp
e^{-i\vec{k}.\vec{x}}
\rsb \frac{d^{3}k}{\lsp2\pi\rsp^{3/2}}
\ee
where, $\varepsilon^{\mu}_{r}\lsp\vec{k}\rsp=\lsp0,
\vec{\varepsilon}_{r}\lsp\vec{k}\rsp\rsp$.
The orthonormality and completeness conditions are given as: 
\be
\varepsilon_{r}\lsp\vec{k}\rsp\varepsilon_{s}\lsp\vec{k}\rsp=\varepsilon_{r
\mu}\lsp\vec{k}\rsp\varepsilon^{\mu}_{s}\lsp\vec{k}\rsp=-\frac{1}{\zeta_{r}}\delta_{r s}, \qquad r=0,...,3,
\label{b1}
\ee
\be
\sum\limits_{r}\zeta_{r}\varepsilon^{\mu}_{r}\lsp\vec{k}\rsp\varepsilon^{\nu}
_{s}\lsp\vec{k}\rsp=-g^{\mu\nu},
\label{b2}
\ee
\be
\zeta_{0}=-1/a^{2}, \qquad \zeta_{1}=\zeta_{2}=\zeta_{3}=1/a^{2}.
\label{b3}
\ee

\noindent (ii) Vector fields of covariant form $A_{\mu}$: The Fourier
decomposition is given as: 
\be
A_{\mu}= \int\varepsilon_{\mu r}\lsp\vec{k}\rsp\lsb b_{r}\lsp\vec{k}\rsp 
A\lsp\eta,k\rsp
e^{i\vec{k}.\vec{x}}+
b_{r}^{\dagger}\lsp\vec{k}\rsp 
A^{*}\lsp\eta,k\rsp
e^{-i\vec{k}.\vec{x}}
\rsb \frac{d^{3}k}{\lsp2\pi\rsp^{3/2}}
\ee
where, $\varepsilon_{\mu r}\lsp\vec{k}\rsp=\lsp0,
\vec{\varepsilon}_{r}\lsp\vec{k}\rsp\rsp$.
The orthonormality and completeness conditions are given as: 
\be
\varepsilon_{r}\lsp\vec{k}\rsp\varepsilon_{s}\lsp\vec{k}\rsp=\varepsilon_{r
\mu}\lsp\vec{k}\rsp\varepsilon^{\mu}_{s}\lsp\vec{k}\rsp=-\frac{1}{\zeta_{r}}\delta_{r s}, \qquad r=0,...,3,
\label{c1}
\ee
\be
\sum\limits_{r}\zeta_{r}\varepsilon_{\mu r}\lsp\vec{k}\rsp\varepsilon
_{\nu s}\lsp\vec{k}\rsp=-g_{\mu\nu},
\label{c2}
\ee
\be
\zeta_{0}=-a^{2}, \qquad \zeta_{1}=\zeta_{2}=\zeta_{3}=a^{2}.
\label{c3}
\ee

\bibliography{vectorpert}
\bibliographystyle{JHEP}
\end{document}